\documentclass[aps,prb,onecolumn,showpacs,twoside,floatfix,superscriptaddress]{revtex4}
\usepackage{graphicx}
\usepackage{graphics}
\usepackage{amsmath}
\usepackage{amssymb}
\usepackage{mathrsfs}
\usepackage{epsf}

\begin{document}

\title{Time dependent current in a nonstationary environment: A microscopic approach}

\author{Satyabrata Bhattacharya}
\affiliation{Department of Chemistry, Bengal Engineering and
Science University, Shibpur, Howrah 711103, India}

\author{Suman Kumar Banik}
\email{skbanik@vt.edu} \affiliation{Department of Biological
Sciences, Virginia Polytechnic Institute and State University,
Blacksburg, VA 24061-0406, USA}

\author{Sudip Chattopadhyay}
\email{sudip_chattopadhyay@rediffmail.com}
\altaffiliation{Corresponding author} \affiliation{Department of
Chemistry, Bengal Engineering and Science University, Shibpur,
Howrah 711103, India}

\author{Jyotipratim Ray Chaudhuri}
\email{jprc_8@yahoo.com} \altaffiliation{Corresponding author}
\affiliation{Department of Physics, Katwa College, Katwa, Burdwan
713130, India}

\date{\today}

\begin{abstract}
Based on a microscopic system reservoir model,where the associated
bath is not in thermal equilibrium, we simulate the nonstationary
Langevin dynamics and obtain the generalized nonstationary
fluctuation dissipation relation (FDR) which asymptotically
reduces to the traditional form. Our Langevin dynamics
incorporates non-Markovian process also, the origin of which lies
in the decaying term of the nonstationary FDR. We then follow the
stochastic dynamics of the Langevin particle based on the
Fokker-Planck-Smoluchowski description, in ratchet potential to
obtain the steady and time dependent current in an analytic form.
We also examine the influence of initial excitation and subsequent
relaxation of bath modes on the transport of the Langevin particle
to show that the nonequilibrium nature of the bath leads to both
strong non-exponential dynamics as well as nonstationary current.
\end{abstract}


\maketitle

\section{Introduction}

The traditional theory of noise induced transport deals with a
Langevin equation describing the motion of a model Brownian
particle in an external periodic potential, spatially symmetric or
asymmetric. \cite{ratchet,rmp,rev1,rev2,rev3} The nature of
asymmetry of the external force field, in which the Brownian
particle is moving, is crucial in generating biased directed
motion. While moving in a symmetric potential, the Brownian
particle is unable to generate motion in a preferred direction due
to the detailed balance principle which can be broken easily by
applying an external time dependent perturbation, either
deterministic or random. The correlation time of the external
perturbation needs to be greater than the correlation time of the
fluctuations which the system experiences from its immediate
surroundings, the heat bath. A general approach in this direction
involves the application of a time periodic deterministic field or
the application of a colored noise to the system of interest
\cite{ratchet,rev1,rev2}. Adopting a different approach, one can
create directed motion by putting the Brownian particle in a
biased asymmetric periodic potential from the very beginning. The
spatial bias in the potential is able to overcome the detailed
balance principle and hence can generate motion in a preferred
direction. \cite{rmp,rev1,rev2}

The theory of directed motion has gained wide interdisciplinary
attention to model the phenomena of noise induced transport, where
the interplay of fluctuations and non-linearity of the system
plays an important role. \cite{ratchet,rmp,rev1,rev2,rev3}
Exploitation of the nonequilibrium fluctuations present in the
medium helps to generate directed motion of the Brownian particle.
Presence of spatial anisotropy in the potential together with
nonequilibrium perturbation enables one to extract useful work
from random fluctuations without violating the second law of
thermodynamics. \cite{ratchet,rev1} This leads to its wide
applicability in explaining the mechanism of molecular motors,
\cite{rmp,rev1,howard} tunneling in a Josephson junction,
\cite{rev1} rotation of dipoles in a constant field, \cite{rev1}
phase locked loop, \cite{rev1} directed transport in photovoltaic
and photoreflective materials \cite{photo} and the efficiency of
tiny molecular machine in a highly stochastic environment.
\cite{rev2,rev3,tiny} Motor proteins like kinesins,dyenins and
myosins are versatile biomolecular shuttle cargo encapsulated in
vesicles and are present in the different parts of the cell. In
living cells, transport occurs via the cytoskeletal filaments and
motor proteins. \cite{howard,cell} Motor proteins are also
important ingredients of the mechanism of muscle contraction and
cell division. \cite{cell} The search for physical principles that
enable such tiny molecular machines to function efficiently in a
highly Brownian regime and construction of artificial molecular
rotors which produce controlled directional motion mimicking
molecular motor proteins \cite{motor} are the subject of ongoing
interest.

During the last two decades, several theoretical models have been
proposed using the idea of a Brownian particle moving in a ratchet
potential \cite{ratchet,rev1,rev2,rmp} to explain the transport
mechanism under various nonequilibrium situations. The ratchet
model and its many variants like rocking ratchet, \cite{ratchet}
diffusion ratchet, \cite{ref14} correlation ratchet, \cite{ref15}
flashing ratchet, \cite{ref16} etc., have found wide attention in
recent days. \cite{rev1} To get a unidirectional current, either
spatially asymmetric periodic potentials or time asymmetric
external forces are necessary in these models. In explaining the
above mentioned directional transport phenomena, most of the
theoretical approaches adopt phenomenological models. The first
self consistent microscopic attempt was made by Millonas
\cite{millonas} in the context of construction of a Maxwell's
demon like information engine that extracts work from a heat bath.
In this microscopic construction, the Hamiltonian for the whole
system includes a subsystem, a thermal bath and a nonequilibrium
bath that represents an information source or sink.
\cite{millonas}

In this article, we consider a simple variant of the system
reservoir hamiltonian \cite{millonas} to model the directional
transport processes where the associated bath is in a
nonequilibrium state. The model incorporates some of the features
of Langevin dynamics with a fluctuating barrier \cite{ref18} and
the kinetics due to space dependent friction along with the
presence of local hot spots. \cite{landauer,porto,jrc} Since the
theories of transport processes traditionally deal with stationary
bath, the nonstationary transport processes have remained largely
overlooked so far. We specifically address this issue and examine
the influence of initial excitation and subsequent relaxation of
bath modes \cite{millonas,jrc1,jrc2,rig,popov,bolivar} on the
transport of system particle. We show that relaxation of the
nonequilibrium bath modes may result in strong non-exponential
kinetics and nonstationary current. The physical situation that
has been addressed is that at $t=0_-$, the time just before the
system and the bath are subjected to an external excitation, the
system is appropriately thermalized. At $t=0$, the excitation is
switched on and the bath is thrown into a nonstationary state
which behaves as a nonequilibrium reservoir. We follow the
stochastic dynamics of the system mode after $t>0$. The separation
of the time scales of the fluctuations of the nonequilibrium bath
and the thermal bath to which it relaxes, is such that the former
effectively remains stationary on the fast correlation of the
thermal noise. \cite{jrc1}

The organization of the paper is as follows: We discuss in Sec.II
a microscopic model necessary to compute the transient transport
process where the system in question is not initially thermalized
and the associated bath is thrown into a nonequilibrium and
nonstationary situation by sudden initial excitation of some of
the bath modes. Appropriate elimination of the reservoir degrees
of freedom leads to a non-Markovian Langevin equation,
stochasticity being contributed by both additive thermal noise and
the multiplicative noise due to relaxing nonequilibrium modes. In
Sec.III, following the prescription of Ref.\onlinecite{sancho},
the Fokker-Planck description is provided in position space which
is valid for state dependent dissipation. We then derive the time
dependent solution of the associated Smoluchowski equation for
probability density function. As an application of our
development, in Sec.IV, we consider the motion of a Langevin
particle in a periodic ratchet potential and obtain the stationary
and time dependent average velocity of the Langevin particle and
show that for symmetric periodic potential, the direction of
average velocity depends on the initial excitation of intermediate
bath modes. Summarizing remarks are presented in Sec.V.

\section{The background and the model}

To make the paper self contained, we first discuss the essential
features of the traditional theory of system reservoir dynamics in
this section and then describe the model we adopt in the present
work. This shows how our model deviates from the usual system
reservoir theory and brings up the new features of our model.

\subsection{The traditional system reservoir model}

In the traditional system reservoir model, \cite{micro,open,ref2}
the reservoir is assumed to be in equilibrium at $t=0$ in the
presence of the system, and the appropriate distribution of the
initial state of the heat bath is governed by the Hamiltonian
\begin{eqnarray}\label{eq1}
H_B + H_{SB}  = \sum_{\nu} \left [ \frac{p^2_{\nu} }{2m_{\nu}} +
\frac{m_{\nu} \omega_{\nu}^2}{2} \left ( q_{\nu} - \frac{g_{\nu}
x}{m_{\nu} \omega_{\nu}^2} \right )^2 \right ],
\end{eqnarray}

\noindent which includes the static interaction part, $H_{SB}$,
between the system and the reservoir. The total Hamiltonian of the
system plus bath is then usually written as
\begin{eqnarray}\label{eq2}
H = \frac{p^2}{2} + V(x) + H_B + H_{SB}.
\end{eqnarray}

\noindent In Eqs.(\ref{eq1}-\ref{eq2}), the system (mass weighted)
is described by the coordinate $x$ and the conjugate momentum $p$,
and the heat bath, composed of a set of linear harmonic
oscillators, by the coordinate $q_{\nu}$ and the conjugate momenta
$p_{\nu}$, $\nu = 1,2 \cdots N $. $m_{\nu}$ is the mass of the
$\nu$-th oscillator and $\omega_{\nu}$, the corresponding
frequency. The system bath interaction is generally taken to be
linear in both the system and the bath coordinates through the
coupling constant $g_{\nu}$. $V(x)$ represents the external force
field in which the Brownian particle is executing random motion.
The bath is assumed to be in thermal equilibrium at temperature
$T$ and the initial distribution is considered to be a canonical
one \cite{micro,open,ref2}
\begin{eqnarray}\label{eq3}
W[{\bf q(0),p(0)}] = \frac{1}{Z} \exp \left (
-\frac{H_B+H_{SB}}{k_BT} \right ),
\end{eqnarray}

\noindent where, $Z$ is the normalization constant and $k_B$ is
the Boltzmann constant. To derive the dynamical equations for the
system in terms of $x$ and $p$, one usually eliminates the bath
degrees of freedom from the equations of motion of the system
variable \cite{micro} and obtains,
\begin{eqnarray}\label{eq4}
\dot{x} &=& p, \nonumber \\
\dot{p} &=& - V^{'}(x) - \int_0^t d\tau \gamma (t-\tau)p(\tau) + \xi (t),
\end{eqnarray}

\noindent where $\gamma(t)$ is the memory kernel
\begin{eqnarray*}
\gamma(t) =  \sum_{\nu} \frac{g_{\nu}^2} {m_{\nu} \omega_{\nu}^2 }
\cos\omega_{\nu}t,
\end{eqnarray*}

\noindent and $\xi(t)$, the forcing function
\begin{eqnarray}\label{eq55}
\xi(t) = \sum_{\nu} g_{\nu} \left [ \left \{ q_{\nu}(0) -
\frac{g_{\nu}}{m_{\nu} \omega_{\nu}^2} x(0) \right \}
\cos \omega_{\nu} t + \frac{p_{\nu}(0)} {m_{\nu} \omega_{\nu}} \sin
\omega_{\nu} t \right ].
\end{eqnarray}

\noindent Having chosen a distribution for the initial state of the
bath, given by Eq.(\ref{eq3}), the fluctuating force $\xi(t)$ becomes
zero centered, and the correlation function of $\xi(t)$ gives the
celebrated fluctuation dissipation relation (FDR) \cite{micro,
open,ref2}
\begin{eqnarray}\label{eq5}
\langle \xi(t) \xi(t^{'})\rangle = k_B T \gamma(t-t').
\end{eqnarray}

\noindent To complete the identification of Eq.(\ref{eq4}) as a
generalized Langevin equation, one must establish the conditions
on the coupling coefficients $g_{\nu}$, on the bath frequency
$\omega_{\nu}$ and on the number $N$ of the bath oscillators which
ensure that $\gamma(t)$ is indeed dissipative. Sufficient
conditions for $\gamma(t)$ to be dissipative are that it is
positive definite and decreases monotonically with time. Both the
conditions are achieved if $N \rightarrow\infty$ and if
$g_\nu/m_\nu \omega_\nu^2$ and $\omega_{\nu}$ are sufficiently
smooth functions of $\nu$. \cite{ref3} As $N \rightarrow\infty$,
one replaces the sum by an integral over $\omega$ weighted by a
density of states $D(\omega)$ to get
\begin{eqnarray}\label{eq6}
\gamma(t) = \int d\omega  D(\omega) c(\omega) \cos(\omega t),
\end{eqnarray}

\noindent with  $(g_\nu/m_\nu \omega_\nu^2) \rightarrow c(\omega)$.
For
\begin{eqnarray}\label{eq7}
D(\omega) c(\omega) = \frac{\gamma/\tau_c}{1+ \tau_c^2 \omega^2},
\end{eqnarray}

\noindent which can be achieved by a variety of combinations of
the density of states $D(\omega)$ and the coupling function
$c(\omega)$, and which broadly resembles the behavior of
hydrodynamic model in a macroscopic system, \cite{open} the
dissipation kernel $\gamma(t)$ becomes
\begin{eqnarray}\label{eq8}
\gamma(t) = \frac{\gamma}{\tau_c} \exp (-|t|/\tau_c).
\end{eqnarray}

\noindent $\tau_c$ in the above expression is the cut-off frequency
and is characterized as the correlation time of the bath. In the
limit $\tau_c \rightarrow 0$, $\gamma(t) \rightarrow 2 \gamma \delta(t)$
and one obtains the traditional Langevin equation in the Markovian domain
\begin{eqnarray}\label{eq9}
\dot{x} &=& p, \nonumber \\
\dot{p} & = & -V^{'}(x) - \gamma p + \xi(t),
\end{eqnarray}

\noindent where, $\langle \xi(t) \rangle = 0 $ and $\langle \xi(t)
\xi(t')\rangle = 2 \gamma k_B T \delta (t-t')$. If one considers
the dynamics of the Brownian particle in a periodic potential $V(x)
= V(x+L)$, whose spatial symmetry can be broken by an external load
(force) thereby creating a biased force field, then the system's
dynamics is governed by
\begin{eqnarray}\label{eq10}
\dot{x} &=& p, \nonumber \\
\dot{p} & = & -V^{'}(x) - \gamma p + \xi(t) +F,
\end{eqnarray}

\noindent where $F$ is the external force. The sum of the periodic
potential $V(x)$ and the potential $-Fx$ due to the external force
$F$, i.e., $U(x) = V(x) -Fx$, is a corrugated plane whose average
slope (a measurement of the bias) is determined by the external
force $F$. \cite{ref4}

Eq.(\ref{eq10}) is the standard Langevin equation of a particle
moving in an external potential under an external load force and
is Markovian in nature. In addition to that, the dissipation term
$\gamma$ is constant due to the linear system reservoir coupling
$g_\nu$ and the noise term $\xi(t)$ is Gaussian, additive in
nature reflecting the Markovian kinetics of the Brownian particle.
In the following subsection, we show how this Markovian kinetics
changes to a non-Markovian one due to the sudden excitation of the
few bath modes and splits the noise term $\xi (t)$ into two parts.

\subsection{The nonstationary system reservoir model}

We consider a Brownian particle of unit mass, described by the
coordinate $x$ and the conjugate momentum $p$, moving in a
periodic potential of periodicity $L$, i.e. $V(x+L) = V(x)$. It is
acted upon by an external force $F$, which for the present study
is assumed to be constant and time independent. The system mode is
coupled to a set of relaxing modes considered as a semi-infinite
dimensional system ($\{q_k\}$-subsystem) which effectively
constitutes a nonequilibrium bath. \cite{millonas,jrc1,popov}
These $\{q_k\}$ modes are in contact with a thermally equilibrated
reservoir. Both the reservoirs are composed of two sets of
harmonic oscillators of unit mass characterized by the frequency
sets $\{ \omega_k\}$ and $\{ \Omega_j\}$ for the nonequilibrium
and the equilibrium bath respectively. The system reservoir
combination evolves under the total Hamiltonian
\begin{eqnarray}\label{eq11}
H = \frac{p^2}{2} + V(x) - Fx + \frac{1}{2} \sum_j (P_j^2 +
\Omega_j^2 Q_j^2)
+ \frac{1}{2} \sum_k (p_k^2 + \omega_k^2 q_k^2) -x\sum_k \kappa_j
Q_j - g(x) \sum_k q_k - \sum_{j,k} \alpha_{jk} q_k Q_j.
\end{eqnarray}
\noindent In Eq.(\ref{eq11}), $\kappa_j$ is the coupling constant
describing the coupling of the system with the equilibrium bath
modes and $g(x)$ is the coupling function. The term $g(x) \sum_j
q_j$ indicates the coupling of the nonequilibrium bath to the
system and the last term describes the coupling between the
nonequilibrium bath and the thermal bath with coupling constant
$\alpha_{jk}$. The equilibrium bath is assumed to be in thermal
equilibrium at a temperature $T$ and the initial distribution of
equilibrium bath variables are assumed to Gaussian. The form of
the nonequilibrium bath, that of a set of phonons or photons, is
chosen for both simplicity and because of its generic relationship
to many condensed matter type systems. \cite{ref2}

Eliminating the equilibrium bath variables $\{Q_j,P_j\}$ in the
traditional way, \cite{micro,open,ref2} one may show that the
nonequilibrium bath modes obey the dynamic equations
\begin{eqnarray} \label{eq12}
\dot{q_k} &=& p_k, \nonumber \\
\dot{p_k} &=& -\gamma p_k -\omega_k^2 q_k - g(x) + \eta_k(t).
\end{eqnarray}

\noindent Eq.(\ref{eq12}) takes into account the average
dissipation $\gamma$ of the nonequilibrium reservoir modes $q_k$
due to their coupling to the thermal reservoir which induces
fluctuations $\eta_k(t)$ characterized by the usual FDR
$\langle\eta_k(t)\eta_k(0)\rangle = 2 \gamma k_B T\delta(t)$.
\cite{micro,jrc1} In general, $\langle\eta_k(t)\rangle$ being a
non-zero constant quantity which, without loss of any generality,
may be chosen as zero by shifting the origin of our coordinate
system as we are dealing with a periodic potential. In passing we
mention that in deriving Eq.(\ref{eq12}) from Eq.(\ref{eq11}), the
cross terms for $\sum_j \gamma_{kj} q_j$ have been neglected.

Proceeding similarly to eliminate the thermal reservoir variables
from the equations of motion of the system, we obtain
\begin{eqnarray}\label{eq13}
\dot{x}&=& p, \nonumber \\
\dot{p} &=& -\gamma_e p - V^{'}(x) + F +\xi_e(t) + g^{'}(x) \sum_k q_k.
\end{eqnarray}

\noindent where $\gamma_e$ refers to the dissipation coefficient
of the system mode due to its direct coupling to the thermal bath
providing fluctuations $\xi_e(t)$. The statistical properties of
$\xi_e(t)$ are $\langle \xi_e(t)\rangle = 0$ and $\langle \xi_e(t)
\xi_e(t^{'})\rangle = 2 \gamma_e k_B T \delta(t -t')$. Comparing
with Eq.(\ref{eq10}), it is easy to see that the dissipation term
$\gamma_e$ and the noise term $\xi_e (t)$ are basically $\gamma$
and $\xi(t)$, respectively, that arise due to the direct linear
system reservoir coupling. Now making use of the formal solution
of Eq.(\ref{eq12}) which takes into account the relaxation of the
nonequilibrium modes, and integrating over the nonequilibrium bath
with a Debye type frequency distribution of the form \cite{jrc1}
\begin{eqnarray}
\rho (\omega) &=& \frac {3 \omega^2}{2\omega_c^3} \text{ for }
|\omega|\leq\omega_c, \nonumber \\
&=&  0  \text{ for } |\omega|>\omega_c,
\end{eqnarray}

\noindent where $\omega_c$ is the high frequency Debye cut-off,
one finally obtains the following Langevin equation for the system
mode, from Eq.(\ref{eq13}) as
\begin{eqnarray}\label{eq14}
\dot{x} &=& p, \nonumber \\
\dot{p} &=& - \Gamma(x) p - \widetilde{V}'(x)
+F +\xi_e(t)  +  g^{'}(x) \xi_n(t).
\end{eqnarray}

\noindent In the above Eq.(\ref{eq14})
\begin{eqnarray}\label{eq15}
\Gamma(x) = \gamma_e + \gamma_n [g^{'}(x)]^2,
\end{eqnarray}

\noindent is the  state dependent dissipation constant comprising
of $\gamma_n$ and $\gamma_{e}$. $\xi_n$ refers to the fluctuations
of the nonequilibrium bath modes which effectively cause a damping
of the system mode. This damping is also state dependent due to
the nonlinear coupling function $g(x)$ as is given by $\gamma_n
[g^{'}(x)]^2$ . In Eq.(\ref{eq14}), the potential $V(x)$ in which
the particle  moves has been modified to
\begin{eqnarray}\label{eq16}
\widetilde{V}(x) = V(x) - \frac{\omega_c}{\pi} \gamma_n g^2(x).
\end{eqnarray}

\noindent The fluctuations $\xi_n(t)$ due to the presence of
nonequilibrium bath is also assumed to be Gaussian with zero mean
$\langle\xi_n(t)\rangle = 0$. Also, the essential properties of
$\xi_n(t)$ explicitly depend on the nonequilibrium state of the
intermediate oscillator modes $\{q_j\}$ through $u(\omega,t)$, the
energy density distribution function at time $t$ in terms of the
following FDR for the nonequilibrium bath \cite{jrc1}
\begin{eqnarray}\label{eq17}
u(\omega,t) = \frac{1}{4\gamma_n} \int_{-\infty}^{+\infty} d\tau
\langle \xi_n(t) \xi_n(t+\tau)\rangle e^{i\omega\tau}
= \frac{1}{2} k_BT + e^{-\gamma t/2} \left [ u(\omega,0) - \frac{1}{2} k_BT \right].
\end{eqnarray}

\noindent $[ u(\omega,0) - (k_BT/2) ]$ is a measure of the
departure of energy density from thermal average at $t=0$. The
exponential term $\exp (-\gamma t/2)$ implies that this deviation,
due to the initial excitation, decays asymptotically to zero as
$t\rightarrow \infty$, so that one recovers the usual FDR for the
thermal bath. \cite{jrc1,popov} Eq.(\ref{eq17}) thus attributes
the nonstationary character of the $\{q_k\}$-subsystem. At this
point it is pertinent to note that the above derivation is based
on the assumption that $\xi_n(t)$ is effectively stationary on the
fast correlation time scale of the equilibrium bath modes. This is
necessary for the systematic separation of the time scales
involved in the dynamics.

Eq.(\ref{eq14}) is the required Langevin equation for the particle
moving in a modified potential $\widetilde{V}(x)$ and is acted
upon by a uniform force $F$. The motion of the particle is damped
by a state dependent friction $\Gamma(x)$. Depending on the
coupling function $g(x)$, both $\widetilde{V}(x)$ and $\Gamma(x)$
are, in general, nonlinear in nature. As a result, the stochastic
differential Eq.(\ref{eq14}) becomes nonlinear. The fluctuating
part in Eq.(\ref{eq14}) is comprised of two quantities;
$\xi_e(t)$, an additive noise due to thermal bath and $\xi_n(t)$,
a multiplicative noise due to nonlinear coupling to the
$\{q_k\}$-subsystem. The Langevin equation (\ref{eq14}) describes
a non-Markovian process as well, where the non-Markovian nature is
characterized by the decaying term in Eq.(\ref{eq17}), describing
the initial nonequilibrium nature of the $\{q_k\}$-subsystem
created by applying sudden excitation at $t=0$. \cite{jrc1,popov}

\section{Stochastic dynamics in the overdamped regime and
the time dependent distribution}

For large dissipation, i.e., in the overdamped limit one usually
eliminates the fast variable $p$ adiabatically by omitting the
inertial term $dp/dt$ from the dynamical equations of motion to
get a simpler description of the system in position space. The
approach of adiabatically eliminating fast variables is valid on a
much slower time scale and is a zero order approximation. For
constant large dissipation, this adiabatic elimination of the fast
variables leads to the correct description of the system's
dynamics. However, in presence of hydrodynamic interactions, i.e.,
when the dissipation is state dependent, the traditional adiabatic
reduction of fast variables does not work properly and gives an
incorrect description of the system's dynamics. For state
dependent dissipation, an alternative approach was proposed in
Ref.\onlinecite{sancho}. Using the method given in
Ref.\onlinecite{sancho}, and using Eq.(\ref{eq14}) one may carry
out a systematic expansion of the relevant variable in powers of
$1/\gamma_e$ by neglecting terms smaller than  $O(1/\gamma_e)$.
Then, by Stratonovich interpretation, it is possible to obtain the
appropriate Langevin equation corresponding to a Fokker-Planck
equation (FPE) in position space. Thus, following
Ref.\onlinecite{sancho}, the formal FPE for the probability
density function (PDF) $P(x,t)$ corresponding to the process
described by Eq.(\ref{eq14}) can be obtained as
\begin{eqnarray}\label{eq18}
\frac{\partial P}{\partial t} & = & \frac{\partial}{\partial x}
\left \{ \frac{\widetilde{V}^{\prime}(x) - F} {\Gamma(x)}  P \right
\} + \gamma_e k_B T \frac{\partial}{\partial x} \left\{
\frac{1}{\Gamma(x)}  \frac{\partial}{\partial x} \frac{1}{\Gamma(x)}
P \right \}
+ \gamma_n k_B T \left( 1 + r e^{- \gamma t/2} \right)
\frac{\partial}{\partial x} \left \{ \frac{g^{'}(x)}{\Gamma(x)}
\frac{\partial}{\partial x} \frac{g^{'}(x)}{\Gamma(x)} P
\right\} \nonumber \\
&& + \gamma_n k_B T \left( 1 + r e^{- \gamma t/2} \right)
\frac{\partial}{\partial x} \left \{
\frac{g^{'}(x)g^{''}(x)}{\Gamma^2(x)} P \right \},
\end{eqnarray}

\noindent where $r = \left \{ [u(\omega \rightarrow 0,0)/2k_BT ]-1
\right \}$ and is a measure of the deviation from equilibrium at
$t=0$. Under the steady state condition (at $t \rightarrow
\infty$), $\partial P/\partial t =0 $ and the stationary
distribution obeys the following relation,
\begin{eqnarray}\label{eq19}
k_B T \frac{d P_S(x)}{dt} + \left ( \widetilde{V}'(x) -F\right) P_S(x)
=0,
\end{eqnarray}

\noindent which has the solution
\begin{eqnarray}\label{eq20}
P_s(x) = N \exp \left[-\frac{1}{k_B T} \int^x \left (
\widetilde{V}^{'}(x^{'}) -F \right) dx^{'} \right],
\end{eqnarray}

\noindent where $N$ is the normalization constant. In Stratonovich
description, the Langevin equation corresponding to the FPE given
by Eq.(\ref{eq19}) is
\begin{eqnarray}\label{eq21}
\dot{x} = - \frac{( \widetilde{V}^{'}(x) -F )}{\Gamma(x)} -
\frac{\widetilde{D}(t) g^{'}(x)g^{''}(x)}{\Gamma^2(x)} +
\frac{1}{\Gamma(x)}
\xi_e(t) + \frac{g^{'}(x)}{\Gamma(x)} \xi_n(t),
\end{eqnarray}

\noindent with $\widetilde{D} = \gamma_n k_B T \left( 1 + r \exp(-
\gamma t/2) \right) $ being the time dependent diffusion constant due
to the relaxation of nonequilibrium bath modes. \cite{jrc1} Let us
consider that the time dependent solution of Eq.(\ref{eq18}) is
given by \cite{jrc2}
\begin{eqnarray}\label{eq22}
P(x,t) = P_S(x) \exp(-\phi(t)),
\end{eqnarray}

\noindent where $\phi $ is a function of time only and
$\lim_{t\rightarrow \infty} \phi(t) =0$. $P_S(x)$ is the steady
state solution of Eq.(\ref{eq18})
\begin{eqnarray}\label{eq23}
&& \frac{d}{dx} \left\{  \frac{( \widetilde{V}^{'}(x) -F
)}{\Gamma(x) } P_S(x) \right\}
+\gamma_e k_B T \frac{d}{dx} \left\{ \frac{1}{\Gamma(x)}
\frac{d}{dx} \frac{1}{\Gamma(x)} P_S(x)\right \}
+ \gamma_n k_B T  \frac{d}{dx} \left \{
\frac{g^{'}(x)}{\Gamma(x)} \frac{d}{dx}
\frac{g^{'}(x)}{\Gamma(x)} P_S(x) \right\} \nonumber \\
&& + \gamma_n k_B T  \frac{d}{dx} \left \{ \frac
{g^{'}(x)g^{''}(x)}{\Gamma^2(x)} P_S(x) \right \} =0.
\end{eqnarray}

\noindent Substitution of Eq.(\ref{eq22}) in Eq.(\ref{eq18})
separates the space and time parts and we have the dynamic
equation for $\phi(t)$
\begin{eqnarray*}
-\frac{d\phi}{dt} \exp(\gamma t/2) ={\rm const}= \alpha \text{
(say)}.
\end{eqnarray*}

\noindent On integration over time we get,
\begin{eqnarray}\label{eq24}
\phi(t) = \frac{2 \alpha}{\gamma} \exp(-\gamma t/2),
\end{eqnarray}

\noindent where $\alpha$ can be determined from the initial
condition. The time dependent solution of Eq.(\ref{eq18}) thus
reads as
\begin{eqnarray}\label{eq25}
P(x,t) = P_S(x) \exp \left [ -\frac{2 \alpha}{\gamma} \exp(-\gamma
t/2)\right].
\end{eqnarray}

\noindent To determine $\alpha$, we now demand that just at the
moment the system (and the non-thermal bath) is subjected to
external excitation at $t=0$, the distribution must coincide with
the usual Boltzmann distribution where the energy term in the
Boltzmann factor, in addition to the usual kinetic and potential
energy terms, contains the initial fluctuation of energy density
$\Delta u[= u(\omega,0) - (k_B T/2)]$. This demands that
\begin{eqnarray}\label{eq26}
\alpha = \frac{\gamma \Delta u}{2 k_BT},
\end{eqnarray}

\noindent $\alpha$ is thus determined in terms of relaxing mode
parameters and fluctuations of the energy density distribution at
$t=0$.

\section{Stationary and Transient Current}

In the over damped limit, the stationary current from
Eq.(\ref{eq23}) can be represented as
\begin{eqnarray}\label{eq27}
J_S = -\frac{1}{\Gamma(x)} \left [ {\widetilde V^{'}}(x) -F + k_BT
\frac{d}{dx}\right]P_S(x).
\end{eqnarray}

\noindent Integrating Eq.(\ref{eq27}) we have the expression for
stationary probability distribution in terms of stationary current
as
\begin{eqnarray}\label{eq28}
P_S(x) = e^{-U(x)}  h(x)
\left [ \frac{P_S(0)}{h(0)} - \frac{J_S \gamma_e} {k_B
T} \int_0^x h(x^{'}) e^{U(x^{'})} dx^{'} \right]
\end{eqnarray}

\noindent where $h(x) = 1+ (\gamma_n/\gamma_e)[g^{'}(x)]^2$,
$\Gamma(x) = \gamma_e h(x)$ and $U(x) = \gamma_e \int_0^x
dx^{\prime} h(x^{'}) [{\widetilde{V}^{'}}(x^{'}) -F]/ k_BT $. We
now consider a symmetric periodic potential with periodicity $L$,
i.e. $V(x) = V(x+L)$ as well as the periodic derivative of
coupling function with the same periodicity as that of the
potential, i.e., $g^{'}(x) = g^{'}(x+L)$. As a consequence of this
choice, $U(x)$ is also a periodic function of $x$ with the period
$L$. If we impose the condition that $P_S(x)$ is bounded for large
enough $x$, it follows from the above mentioned conditions of
periodicity, that $P_S(x+L)= P_S(x)$ i.e. $P_S(x)$ must be
periodic with the same period $L$. \cite{ref4} Now applying the
periodicity condition of $P_S(x)$, we have from Eq.(\ref{eq28})
\begin{eqnarray}\label{eq30}
\frac{P_S(0)}{h(0)}  = J_S \frac{ \gamma_e/k_BT} {1-e^{U(L)}}
\int_0^L h(x) e^{U(x)} dx.
\end{eqnarray}

\noindent Because of the periodicity, we normalize the steady state
PDF in the periodic interval
\begin{eqnarray}\label{eq31}
\int_0^L P_S(x) dx =1,
\end{eqnarray}

\noindent to get
\begin{eqnarray}\label{eq32}
\int_0^L h(x) e^{-U(x)}
\left [ \frac{P_S(0)}{h(0)} - \frac{J_S \gamma_e}{k_BT}
\int_0^x h(x^{'}) e^{U(x^{'})} dx^{'} \right ]dx =
1.
\end{eqnarray}

\noindent Now eliminating $P_S(0)/h(0)$ from Eq.(\ref{eq30}) and
Eq.(\ref{eq32}), one obtains the steady state current
\begin{eqnarray}\label{eq33}
J_S &=& \frac{k_BT}{\gamma_e} \left [ 1 - e^{U(L)} \right ]
\left [\int_0^L h(x) e^{-U(x)} dx \int_0^L   h(x^{'})
e^{U(x^{'})} dx^{'}
 \right. \nonumber \\
&& \left. - \left [ 1 - e^{U(L)} \right ]
\int_0^L \left ( h(x) e^{-U(x)} \int_0^x h(x^{'})
e^{U(x^{'})} dx^{'} \right ) dx \right
]^{-1}.
\end{eqnarray}

\noindent From Eq.(\ref{eq33}) it is clear that in the absence of
any external bias $F$, the steady current vanishes. We thus observe
that there is no occurrence of current for a periodic potential and
for periodic derivative of the coupling function with the
same periodicity for $F=0$. At
the macroscopic level this confirms that there is no generation of
perpetual motion of the second kind, i.e., no violation of second law of
thermodynamics. In passing, we note that in the absence of
$\{q_k\}$-subsystem, i.e., when $\gamma_n =0 $ , Eq.(\ref{eq33})
reduces to the standard form \cite{ref4}
\begin{eqnarray}\label{eq34}
J_S &=& L \gamma_e k_B T \left [ 1 - e^{L F/k_BT} \right ]
\left [\int_0^L e^{V(x)/k_BT} dx \int_0^L e^{-V(x)/k_BT} dx -
\left [1 - e^{-2L F/k_BT} \right ] \right. \nonumber \\
&& \left. \times \left \{ \left ( \int_0^L e^{-V(x)/k_BT} \int_0^x
e^{V(x^{'})/k_BT} dx^{'} \right) dx \right \}
\right]^{-1}.
\end{eqnarray}

\noindent Next, to find the time dependent current $J(x,t)$ we
resort to Eq.(\ref{eq18}) and observe that
\begin{eqnarray}\label{eq35}
J(x,t) & = & -  \frac{\partial}{\partial x} \left \{
\frac{\widetilde
{V}(x^{\prime}) - F} {\Gamma(x)}  P \right \}
+ \gamma_e k_B T   \left\{ \frac{1}{\Gamma(x)}
\frac{\partial}{\partial x} \frac{1}{\Gamma(x)}
P \right \}
+ \gamma_n k_B T \left( 1 + r e^{- \gamma t/2} \right)
 \left \{ \frac{g^{'}(x)}{\Gamma(x)}
\frac{\partial}{\partial x} \frac{g^{'}(x)}{\Gamma(x)} P
\right\} \nonumber \\
&& + \gamma_n k_B T \left( 1 + r e^{- \gamma t/2} \right) \left \{
\frac{g^{'}(x)g^{''}(x)}{\Gamma^2(x)} P \right \}.
\nonumber \\
\end{eqnarray}

\noindent Now substituting Eq.(\ref{eq22}) in Eq.(\ref{eq35}) and
making use of Eq.(\ref{eq23}) we find that $J(x,t)$ can be expressed
in a much simpler form
\begin{eqnarray}\label{eq36}
J(x,t) = J_S e^{-\phi(t)} - {\cal {D}}(t) \frac{1}{\Gamma(x)}
\frac{d}{dx} \frac{1}{\Gamma(x)} [g^{'}(x)]^2 P_S(x),
\end{eqnarray}

\noindent where $P_S(x)$ is the steady state PDF and $J_S$ is the
steady state current given by Eq.(\ref{eq34}) and
\begin{eqnarray}\label{eq37}
{\cal{D}}(t) = r \gamma_n k_B T e^{- \gamma t /2} e^{-\phi(t)}.
\end{eqnarray}

\noindent The steady state current $J_S$ thus can be obtained from
\begin{eqnarray}\label{eq38}
J_S = - \frac{( \widetilde{V}^{'}(x) -F )}{\Gamma(x)} P_S(x) -
\gamma_e k_B T   \frac{1}{\Gamma(x)} \frac{d}{dx}
\frac{1}{\Gamma(x)} P_S(x)
- \gamma_n k_B T  \frac{g^{'}(x)}{\Gamma(x)}
\frac{d}{dx} \frac{g^{'}(x)}{\Gamma(x)} P_S(x)
- \gamma_n k_B T \frac{g^{'}(x)g^{''}(x)}{\Gamma^2(x)} P_S(x),
\end{eqnarray}

\noindent from which we have
\begin{eqnarray}\label{eq39}
\frac{1}{\Gamma(x)} \frac{d}{dx} \frac{1}{\Gamma(x)}
[g^{'}(x)]^2 P_S(x)
= - \frac{J_S} {\gamma_n k_B T } - \frac{( \widetilde{V}^{'}(x)
-F )}{ \gamma_n k_B T \Gamma(x)}
- \frac{\gamma_e} {\gamma_n} \frac{1}{\Gamma(x)} \frac{d}{dx}
\frac{1}{\Gamma(x)} P_S(x).
\end{eqnarray}

\noindent Using Eq.(\ref{eq39}) we then obtain from Eq.(\ref{eq36})
\begin{eqnarray}\label{eq40}
J(x,t) = J_S \left [ e^{-\phi(t)} + \frac{{\cal D}(t) }{\gamma_n
k_B T} \right]
+\frac{{\cal D}(t) }{\gamma_n k_B T} \left [ \frac{(
\widetilde{V}^{'}(x) -F )}{\Gamma(x)} P_S(x)
+ \gamma_e k_B T \frac{1}{\Gamma(x)} \frac{d}{dx}
\frac{1}{\Gamma(x)} P_S(x) \right ].
\end{eqnarray}

\noindent Defining the space dependent part on the RHS of
Eq.(\ref{eq40}) as $M(x)$, we obtain
\begin{eqnarray}\label{eq41}
J(x,t) = J_S \left [ e^{-\phi(t)} + \frac{{\cal D}(t) }{\gamma_n k_B
T}\right] + \frac{{\cal D}(t) }{\gamma_n k_B T} M(x),
\end{eqnarray}

\noindent where
\begin{eqnarray}\label{eq42}
M(x) & = & \left [ \frac{( \widetilde{V}^{'}(x) -F )}{\Gamma(x)} P_S(x)
+ \gamma_e k_B T \frac{1}{\Gamma(x)} \frac{d}{dx}
\frac{1}{\Gamma(x)} P_S(x) \right ]. \nonumber \\
\end{eqnarray}

\noindent From Eq.(\ref{eq41}) we observe that the current
$J(x,t)$ can be written as a sum of two terms. The first term is
space independent and only a function of time. The second term is
product separable in the form of time and space part. As $t
\rightarrow \infty$, RHS of ${\cal {D}}(t) \rightarrow 0$ and
asymptotically $J(x,t)$ reduces to the steady state current $J_S$.
Now using the continuity equation
\begin{eqnarray*}
\frac{\partial P(x,t)}{\partial t} = - \frac{\partial
J(x,t)}{\partial x},
\end{eqnarray*}

\noindent along with $P(x,t) = P_S(x) e^{-\phi(t)}$, we get from
Eq.(\ref{eq41})
\begin{eqnarray}\label{eq42n}
\frac{dM(x)}{dx} = - \frac{\alpha}{r}P_S(x),
\end{eqnarray}

\noindent or equivalently
\begin{eqnarray}\label{eq43}
M(x) = - \frac{\alpha}{r} \int^x P_S(x) dx.
\end{eqnarray}

\noindent As we are dealing with periodic functions, the constant
of integration is chosen to be zero. Now integrating
Eq.(\ref{eq27}) for $P_S(x)$ and using the normalization
condition, Eq.(\ref{eq31}), we have the expression for steady
state PDF as
\begin{eqnarray}\label{eq44}
P_S(x) = e^{- (\widetilde{V}(x) - F x)/k_B T}
\left [ \frac{1+ \frac{J_S}{k_BT} \int_0^L e^{-
(\widetilde{V}(x) - F x)/k_B T} \{ \int_0^x \Gamma(x^{'})
e^{(\widetilde{V}(x^{\prime}) - F x^{\prime})/k_B T} dx^{'} \} dx }
{\int_0^L e^{- (\widetilde{V}(x) - F x)/k_B T} dx}
\right].
\end{eqnarray}

\noindent Using Eq.(\ref{eq44}) along with Eq.(\ref{eq43}) one
obtains from Eq.(\ref{eq41}) the expression for the time dependent
current, $J(x,t)$ as
\begin{eqnarray}\label{eq45}
J(x,t) & = &  J_S \left [ e^{-\phi(t)} + r e^{-\gamma t/2} \right]
- \frac{\alpha e^{-\gamma t/2} \int^x dx^{'}  e^{-
({\widetilde{V}}(x^{'}) - F x^{\prime})/k_B T} }{\int_0^L e^{-
({\widetilde V}(x) - F x)/k_B T} dx}
\left [ 1+ \frac{J_S}{k_BT} \int_0^L e^{-
({\widetilde{V}}(x^{''}) - F x^{''})/k_B T} \right. \nonumber
\\
&& \left. \times \left \{ \int_0^{x^{'}} \Gamma(x^{'''}) \{ e^{
({\widetilde{V}}(x^{'''}) - Fx^{'''})/k_B T} dx^{^{'''}} \}dx^{''}
\right \} \right ],
\end{eqnarray}

\noindent where $J_S$ is given by Eq.(\ref{eq33}). Since the
potential possesses spatial periodicity, one has $J(x,t) =
J(x+L,t)$. Hence the net time dependent current is given by
\begin{eqnarray}\label{eq46}
j(t) = \frac{1}{L} \int_0^L J(x,t) dx.
\end{eqnarray}

\noindent It should be noted that for symmetric potential with
$F=0$, $J_S=0$. But, in our development, transient current exists
and the direction of current depends on the sign of $\alpha$. What
is immediately apparent is that for symmetric potential, the sign
of $\Delta u [= u(\omega, 0) - (1/2k_BT)]$ determines the
direction of initial current
\begin{eqnarray}\label{eq47}
j(t) = - \frac{\alpha e^{-\gamma t /2}}{L} \frac { \int_0^L dx
\int^x dx^{'} \exp[-{\widetilde{V}}(x^{'})/k_BT ]}  { \int_0^L dx
\exp[- {\widetilde{V}} (x)/k_BT ] }.
\end{eqnarray}

\noindent It is also clear from Eq.({\ref{eq45}) that the time
dependent current reduces to the steady state current, $J_S$ in
the asymptotic limit. The presence of the term $\exp[-\phi(t)]$ in
the expression of $J(x,t)$ makes the transient current strongly
non-exponential in nature. The transient behavior of growth or
decay of charge and current in $L-R$, $C-R$ or $L-C-R$ circuit is
important in construction of many electrical and electronic
devices where there is the mechanism of storage of energy. In
construction of molecular motor or nano-switch, the transient
behavior of the devices may be worth studying. In our development,
the preparation of intermediate relaxing bath plays a key role to
generate the time dependent current. Nevertheless, our methodology
will also be applicable in the case when any arbitrarily prepared
bath is approaching towards equilibrium. In passing, we mention
that the model considered in the present paper may be realized in
a guest-host system embedded in a lattice where the immediate
neighborhood of the guest comprises intermediate oscillatory
modes, while the lattice acts as a thermal bath.

\section{conclusion}
We have hereby proposed a simple microscopic system nonequilibrium
bath model to simulate nonstationary Langevin dynamics. The
nonequilibrium bath is effectively realized in terms of a
semi-infinite dimensional reservoir which is subsequently kept in
contact with a thermal reservoir which allows the non-thermal bath
to relax with characteristic time. The frequency spectrum of the
relaxing bath is assumed to be of Debye type. By an appropriate
separation of time scale, we then construct the Langevin equation
for a particle in which the dissipation is state dependent and the
stochastic forces appearing are both additive and multiplicative.
The underlying stochastic dynamics is found to be nonstationary
and non-Markovian. Based on the strategy of Sancho \emph{et al}.,
\cite{sancho}
we then show that this Langevin equation can be recast into the
form of generalized nonstationary Smoluchowski equation which
reduces to its standard form asymptotically. We then solve the
expression for time dependent PDF. As an immediate application of
our recent development, we consider the dynamics of a Langevin
particle in a ratchet potential and obtain the analytic expression
for both stationary and nonstationary transient average velocity,
which is followed by an immediate observation that in a periodic
potential the direction of nonstationary current depends on the
preparation of the nonequilibrium bath.

\begin{acknowledgments}
We dedicate this article to Professor Jayanta Kr. Bhattacharjee, a
motivating scientist and an inspiring teacher spearheading the
proliferation of Physics as a whole through his skilled
exposition. SB wishes to acknowledge the constant support of his
school authority (Parulia K. K. High School, Burdwan 713513)
towards pursuing his Ph. D. work.
\end{acknowledgments}

\end{document}